\documentclass[11pt,titlepage,dvips]{article}

\usepackage{epsfig,latexsym,amsmath}

\newcommand{\Base}{\baselineskip=12pt} 

\makeatletter
\def\section{\@startsection {section}{1}{\z@}{-3.5ex plus -1ex minus 
+.4ex}{1.5ex plus .2ex}{\large\bf}}
\makeatother

\newcommand{\amp}[1]{\mbox{${\cal A}(|#1\rangle)$}}
\newcommand{\ampp}[1]{\mbox{${\cal A}^{\prime}(|#1\rangle)$}}
\renewcommand{\th}{\mbox{$\mathit{Th}$}}
\newcommand{\eq}{\mbox{$\mathit{Et}$}}
\newcommand{\EC}{\mbox{\rm EC}}
\newcommand{\NC}{\mbox{{\rm NC}}}
\newcommand{\TC}{\mbox{{\rm TC}}}
\newcommand{\QNN}{\mbox{{\rm QNN}}}
\newcommand{\QNNs}{\mbox{{\rm QNNs}}}
\newcommand{\NAND}{\mbox{{\rm NAND}}}

\newtheorem{thm}{Theorem}[section]
\newtheorem{lem}[thm]{Lemma}
\newtheorem{col}[thm]{Corollary}
\newtheorem{defn}[thm]{Definition}
\newtheorem{prop}[thm]{Proposition}

\newcommand{\bthm}{\begin{thm}}
\newcommand{\ethm}{\end{thm}}
\newcommand{\blem}{\begin{lem}}
\newcommand{\elem}{\end{lem} }
\newcommand{\bcol}{\begin{col}}
\newcommand{\ecol}{\end{col} }
\newcommand{\bdefn}{\begin{defn}}
\newcommand{\edefn}{\end{defn} }
\newcommand{\bprop}{\begin{prop}}
\newcommand{\eprop}{\end{prop} }
\newcommand{\bproof}{{\bf Proof. \ }}
\newcommand{\eproof}{\mbox{$\ \ \Box$}\newline }

\begin{document}
\bibliographystyle{plain}

\date{February 23, 2001}


\title{Quantum Neural Networks}
\author{Sanjay Gupta\\
	Department of Computer Science  \\
Virginia Polytechnic Institute and State University \\
7054 Haycock Rd.\\
Falls Church VA 22043-2311\\
Telephone: (703) 538-8373\\
Fax: (703)  538-8348\\
	email: sgupta@vt.edu 
\and R.K.P. Zia \\
Department of Physics \\
Virginia Polytechnic Institute and State University \\
Blacksburg, VA 24061-0435\\
email: rkpzia@vt.edu}
	
\maketitle


\begin{abstract}

This paper initiates the study of quantum 
computing within the constraints of using a polylogarithmic
($O(\log^k n), k\geq 1$) number of qubits and a polylogarithmic
number of computation steps. 
The current research in the literature has focussed on 
using a polynomial number of qubits. 
A new mathematical
model of computation called \emph{Quantum Neural Networks (QNNs)}
is defined,
building on Deutsch's model of quantum computational network. 
The model introduces a nonlinear and irreversible gate, 
similar to the speculative operator defined by Abrams and Lloyd.
The precise dynamics of this operator are defined and 
while giving examples in which nonlinear
Schr\"{o}dinger's equations are applied, we speculate on its possible
implementation.
The many practical problems
associated with the current model of quantum computing
are alleviated in the new model.
It is shown that QNNs of logarithmic size and constant depth
have the same computational power
as threshold circuits, which are used for modeling neural networks. 
QNNs of polylogarithmic size and polylogarithmic depth can solve
the problems in \NC, the class of problems with theoretically
fast parallel solutions. 
Thus, the new model may indeed provide an approach for 
building scalable parallel computers.


\end{abstract}


\section{Introduction}
\label{section:intro}

The concept of quantum computing, based on the quantum mechanical
nature of physical reality, is first stated by Benioff \cite{Ben82}
and Feynman \cite{Fey82}, and formalized by Deutsch
\cite{Deu85}, Bernstein and Vazirani \cite{BV93}, and Yao
\cite{Yao93}.
For background material, the reader is referred to the papers
mentioned above, survey papers \cite{Aha98,Ber97,RP98,Ste98b}, 
books \cite{Gru99,WC98}, and courses
available on the web \cite{Preweb,Vazweb}.
 
Considerable interest has been generated in quantum computing since
Shor \cite{Sho94} showed that numbers can be factored in polynomial
time on a quantum computer. From a practical viewpoint, Shor's result
shows that a working quantum computer can violate the security of
transactions that use the RSA protocol, a standard for secure
transactions on the Internet. From a theoretical viewpoint, the result
seemingly violates the polynomial version of the Church-Turing thesis; it is
generally believed that factoring cannot be done in polynomial time on
a deterministic or probabilistic Turing machine.  What makes
Shor's breakthrough result possible on a quantum Turing machine is that
exponentially many computations can be performed in parallel in one
step and certain quantum steps enable one to extract the desired
information.

Even though simple quantum computers have been built, enormous
practical issues remain for larger-scale machines. 
The problems seem to be exacerbated with more qubits and
more computation steps. 
In this paper, we
initiate the study of quantum 
computing within the constraints of using a polylogarithmic
($O(\log^k n), k\geq 1$) number of qubits and a polylogarithmic
number of computation steps. 
The current research in the literature has focussed on 
using a polynomial number of qubits. 
(Recently, researchers have initiated the study of 
quantum computing using a polynomial number of qubits and a
polylogarithmic number of steps \cite{MN98,Moo99,GHP00,CW00}.)
We define a new mathematical model of computation called
\emph{Quantum Neural Networks (QNNs)}, building on Deutsch's
model of quantum computational network \cite{Deu89}.
The new model introduces a nonlinear, irreversible, and dissipative 
operator, called $D$ gate, similar to the speculative operator introduced
by Abrams and Lloyd \cite{AL98}.
We also define the precise dynamics of this operator and 
while giving examples in which nonlinear
Schr\"{o}dinger's equations are applied, we speculate on the possible
implementation of the $D$ gate.

Within a general framework of size,
depth, and precision complexity, we study the computational 
power of QNNs.
We show that QNNs of logarithmic size and constant depth
have the same computational power
as threshold circuits, which are used for modeling neural networks. 
QNNs of polylogarithmic size and polylogarithmic depth can solve
the problems in \NC, the class of problems that have theoretically
fast parallel solutions.
Thus, the new model subsumes
the computation power of various theoretical models of parallel computation.

We believe that the true advantage of quantum computation
lies in overcoming the communication bottleneck that has plagued 
the implementation of various theoretical models of parallel computation. 
For example, \NC\ circuits elegantly capture the class of problems
that can be theoretically solved fast in parallel using
simple gates. 
While fast implementations of individual gates have been achieved with
semiconductors and millions of gates have been put on a single
chip, we do not have the 
implementation of full \NC\ circuits
because of the communication and synchronization
costs involved in wiring a polynomial number of gates.
We believe that this
hurdle can be overcome using the nonlocal interactions present in
quantum systems --- there is no need to explicitly wire the 
entangled units and the synchronization is instantaneous.
This advantage is manifest in the standard unitary operator,
where operations on one qubit can affect probability amplitudes
on all the qubits, without requiring explicit physical 
connections and a global clock.
Thus, the new model has the potential to overcome the practical
problems associated with both quantum computing as well as classical
parallel computing. 

The paper addresses three categories of researchers: complexity
theory, neural networks, and quantum computing.
For complexity theorists, the paper shows that the 
$2^{\log^{O(1)}n}$ bounds on threshold circuits obtained
in various results such as \cite{AH94} is not necessarily
infeasible. (Polynomial time and space is generally accepted
as defining feasible bounds.)
For neural network researchers, the paper mathematically proves
that threshold circuits used for modeling neural networks have
the same computation power as the equality threshold circuits,
introduced in this paper.  
The new class has certain algebraic advantages and it may
indeed be beneficial to redo the theory of neural networks
under this model.
Finally, for quantum computing researchers, the paper shows that
with quantum computing we can build scalable parallel 
computers, beyond the current digital technology, under
very tight constraints. 
Of course, this depends on the physical realization of the inherently
nonlinear gate introduced in the paper and thus it is worthwhile
exploring its implementation.
Furthermore, we do not need to limit ourselves to the fundamental 
linear models of systems to obtain useful devices; we should explore
systems at a different level of abstractions that may not be
defined by linear equations.

This paper is organized as follows: Section \ref{section:problems}
summarizes the current model of quantum computing,
together with the associated problems, 
including some new issues not discussed in the literature.
Section \ref{section:new} first
defines a framework for new quantum computing models, 
based on the problems discussed in Section \ref{section:problems}. 
The $D$ gate is introduced within this framework. 
(This is the only speculative part of the paper; that is,
we do not know how to implement the $D$ gate.)
QNNs are defined to be Deutsch's quantum computational 
network \cite{Deu89} together with the $D$ gate.

Section \ref{section:power} quantifies the
computational power of the new model. 
We prove that equality threshold
circuits have essentially the same computational power as the standard class
of threshold circuits used for modeling neural networks. 
Then QNNs are shown to have the same computational power
as equality threshold gates.
Section \ref{section:Dgate} defines the precise dynamics of the 
$D$ gate.
Section \ref{section:nonlinear} identifies some of the many nonlinearities in 
quantum systems.
Finally, Section \ref{section:future} delineates many interesting 
research directions suggested by this paper.


\section{Current Model of Quantum Computing}
\label{section:problems}

\begin{figure}
\begin{center}
\input{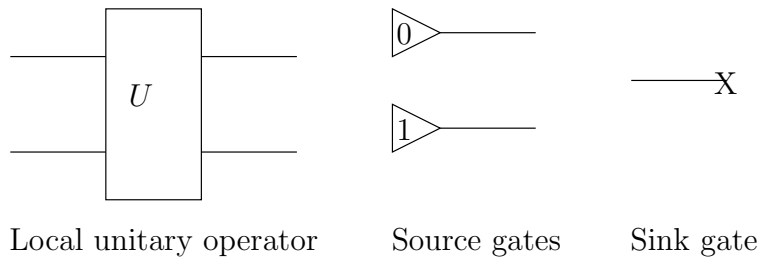}
\caption{Components of Deutsch's quantum computational network model}
\label{fig:QCN}
\end{center}
\end{figure}

There are two equivalent models for quantum computing, quantum
Turing machines \cite{Deu85,BV93} based on reversible Turing machines
\cite{Ben73,Ben89} and quantum computational network \cite{Deu89}. Our 
model builds on the latter, so we briefly review it here.
The basic unit in quantum computation is a \emph{qubit}, a superposition of
two independent states $|0\rangle$ and $|1\rangle$, denoted 
$\alpha_0 |0\rangle + \alpha_1 |1\rangle$, where
$\alpha_0,\alpha_1$ are complex numbers such that
$|\alpha_0|^2 + |\alpha_1|^2 =1$. 
A system with $n$ qubits is described using $2^n$ 
independent states $|i\rangle, 0 \leq i \leq 2^n-1$, 
each associated with \emph{probability amplitude} $\alpha_i$,
a complex number,
as follows: $\sum_{i=0}^{2^n-1} \alpha_i |i\rangle$, where
$\sum_{i=0}^{2^n-1} |\alpha_i|^2 =1$. 
The direction of $\alpha_i$ on the 
complex plane is called the \emph{phase} of state $|i\rangle$
and the absolute value $|\alpha_i|$ is called the \emph{amplitude}
of state $|i\rangle$.

The computation unit in Deutsch's model consists of \emph{quantum gates}
whose inputs and outputs are qubits.
A gate can perform
any local unitary operation on the inputs.  It has been shown that 
one-qubit gates together with two-qubit controlled NOT gates are universal
\cite{BBC95}. 

The quantum gates are interconnected by \emph{wires}.  
A quantum computational network is a computing machine consisting of 
quantum gates with synchronized steps. 
By convention, the computation proceeds from left to right.
The outputs of some of the
gates are connected to the inputs of others. 
Some of the inputs are used as the input to the network. 
Other inputs are connected to
\emph{source} gates for $0$ and $1$ qubits. Some of the outputs
are connected to \emph{sink} gates, where the arriving qubits
are discarded. 
The essential ingredients of the model are summarized
in Figure \ref{fig:QCN}. 
An output qubit can be measured along state
$|0\rangle$ or $|1\rangle$, and is observed based on the probability
amplitudes associated with the qubit. 

Even though simple quantum computers have been built,
enormous practical issues remain for larger-scale
machines.  Landauer \cite{Lan95} exposes three main problems: {\em
decoherence}, {\em localization}, and {\em manufacturing defects.}
Decoherence is the process by which a quantum system decays to a
classical state through interaction with the environment.  In the best
case, coherence is maintained for some $10^{4}$ seconds, and, in the
worst case, for about $10^{-10}$ seconds for single qubits.  Some
decoherence models show the coherence time declining exponentially as
the number of qubits increases \cite{Unr95}.  Furthermore, the
physical media that allow fast operations are also the ones with short
coherence times.

The computation may also suffer from localization, that is, from
reflection of the computational trajectory, causing the computation to
turn around.  Landauer \cite{Lan95} points out that this problem is
largely ignored by the research community.  The combination of
decoherence and localization makes the physical realization of quantum
computation particularly difficult. On the one hand, we need to isolate a
quantum computing system from the environment to avoid decoherence,
and on the other hand, we need to control it externally to compel it to run
forward to avoid reflection. Finally, minor manufacturing defects can
engender major errors in the computations.

Introduction of the techniques of error-correcting codes 
and fault-tolerant computation 
to quantum computation has generated considerable optimism for building
quantum computers, because these techniques can alleviate the problems of
decoherence and manufacturing defects.
Though this line of research is elegant and
exciting, the codes correct only local errors. For example, one qubit
can be encoded into the nonlocal interactions among three qubits to
correct one qubit errors. However, in principle, any (nonlocal) unitary
operator can be applied to all the qubits.  These nonlocal errors
easily subvert the error-correcting codes. Also, nonlocal interactions
provide the exponential speed-ups in quantum computing. 
The hope that nature might allow computational
speed-ups via nonlocal interactions, while errors are constrained to
occur only locally, seems unavailing. For an excellent exposition on
error-correcting codes and fault-tolerant quantum computing, the
reader is referred to \cite{Pre97}.

It has been shown that one-qubit gates together with
two-qubit controlled NOT gates are universal \cite{BBC95}; that is,
any $2^n \times 2^n$ unitary operator can be decomposed
into a polynomial number of one and two qubit operators. 
However, in general,
any error operator can be applied in one step that cannot
even be detected without observing all the involved qubits. Having the 
ability to operate on many qubits does not solve the problem,
for error-correcting codes for $k$ qubit errors can be subverted
by a $(k+1)$-qubit error operator. Eventually, construction of a
$2^n \times 2^n$ operator will itself be more time consuming than
the actual computation.

There are some additional difficulties with computing using a
polynomial number of qubits for a polynomial number of steps that are
not discussed in the literature.  For example, if $n=1000$ and an
$O(n^2)$ quantum algorithm is used, we need one million \emph{uniquely
identifiable} but \emph{identical} carriers of quantum information.
Clearly, the carriers need to be uniquely identifiable because we are
not using their statistical properties, but encoding $2^{O(n^2)}$
computations in their interactions. However, the carriers need to be
absolutely identical for the following reason.  In describing the
Hamiltonian for the whole system, there is a phase oscillation
associated with each carrier. 
If all carriers have the same frequency,
it does not affect the computation, which essentially changes the
state relative to the global oscillation.  
But each qubit is likely to
be encoded in carriers with a much larger state space, and even slight
frequency differences can result in substantial errors over a
polynomial number of steps. The task of preparing one million
absolutely identical carriers, while exploiting the $2^{10^6}$
interactions, most of which are nonlocal, for speeding-up computation
appears insurmountable.

In conclusion, controlling a polynomial number of entangled qubits for
a polynomial number of steps, while compelling the computation
forward, seems hard even with the help of error-correcting codes. 
To address the above problems, we initiate the study of quantum computation
under the constraints of a polylogarithmic number of qubits 
and a polylogarithmic number of steps.


\section{Quantum Neural Networks (QNNs)}
\label{section:new}

In the previous section we saw that the problems with the current
model get worse as the number of qubits and computation steps increase.
Therefore, we make the following premises for a new
model.

\begin{itemize}
\item The computation should be achieved within a few
($O(\log^k n), k\geq 1$) steps.

\item (Optional) The computation should use only a few
($O(\log^k n), k\geq 1$) qubits.

\item There should be irreversible synchronization points
in the computation to avoid the problem of localization.
\end{itemize}

The second point is optional because a single particle,
in principle, can encode an infinite amount of discrete information,
for example, in spins $-s/2,-(s-1)/2,\ldots,s/2$ for any number $s$.
Thus, 
many qubits can be encoded within a single particle, and hence,
it may be worthwhile exploring models that use many qubits
encoded on a few carriers.
However, in this paper we observe all the above premises. 
Note that within the framework of our premises, 
quantum computation does not seem to violate the polynomial
version of the Church-Turing thesis.

The next question is what operations should we have besides
the standard unitary operator. In particular, what irreversible,
and hence, nonlinear operators can be exploited for computation. 
In its present formulation, quantum mechanics for an isolated
system, or one interacting with \emph{classical} external potentials, is
linear in the state vector. However, if full interactions with an
environment are taken into account, then their effects can be found by
projecting the larger state vector of the combined system into the subspace
of our system alone. 
As a result, the evolution equation for the latter
state vector is not necessarily linear. 
The behavior of a system can also be nonlinear because of the
interactions between the degrees of freedom, as discussed in
Section \ref{section:nonlinear}.

In general, these nonlinear 
operations can be on the phases as well as amplitudes
(or both) associated with a quantum system. The phases in a quantum
system have the following properties.
\begin{itemize}
\item Phases cannot be directly observed.
\item They have an oscillation associated with them.
\item Quantum mechanics can be defined using a density matrix
formulation, which is typically used when incorporating 
phase information is not feasible.
\end{itemize}

\begin{figure}
\begin{center}
\input{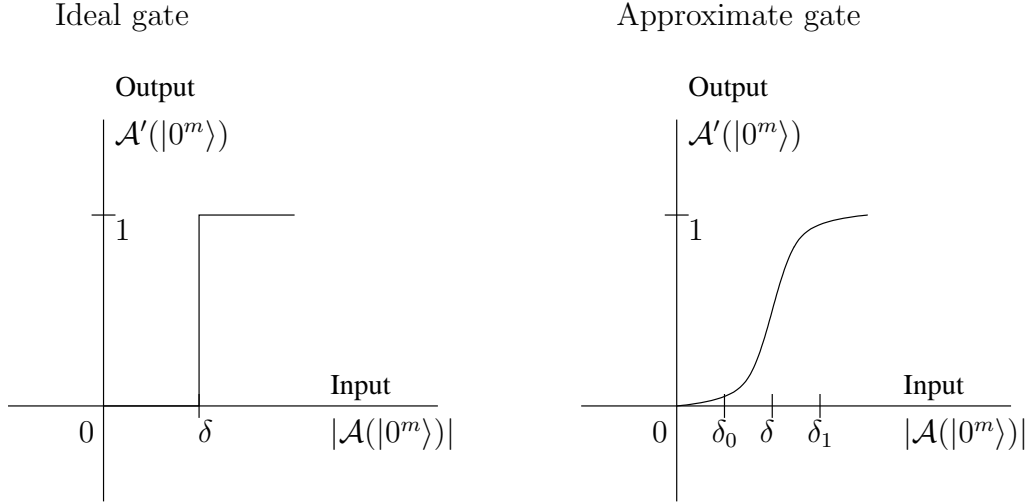}
\caption{Behavior of $D(m,\delta)$ gate}
\label{fig:Dgate}
\end{center}
\end{figure}

Therefore, it seems natural to have nonlinearities on amplitudes only.
Specifically, we introduce a new nonlinear operator
$D$ gate (for dissipative) in Deutsch's quantum computational 
network model \cite{Deu89}.  
The operator  
$D(m,\delta)$, abbreviated $D$ with threshold
$\delta$ (Figure \ref{fig:Dgate}), when applied
to the $m$ qubits in a system with $n\geq m$ qubits behaves as
follows. Let the states of $m$ qubits be represented by 
binary numbers $0$ through $2^m-1$ or, equivalently, 
binary strings $0^{m}$ through $1^{m}$.
Let $\amp{j}$ and $\ampp{j}$ respectively
denote the probability amplitudes before and after the application of
the $D$ operator. Then,
\[ |\amp{0^m}| < \delta \Rightarrow 
\ampp{0^m} = 0\; ,\]
and
\[ |\amp{0^m}| > \delta  \Rightarrow 
\ampp{0^m} = c\; .\]
The value $c$ for probability amplitude denotes some constant used for
encoding $1$. For example, if a system consists of $n$ qubits, then
$\mbox{${\cal A}^{\prime}(|0^m\rangle)$}= 1/\sqrt{2^n}$ is sufficient
for the case $|\mbox{${\cal A}(|0^m\rangle)$}| > \delta$.
In this case, threshold $\delta$ is chosen so that $0< \delta < 1/\sqrt{2^n}$.
(From now on, we'll simply use $1$ instead of $c$.)
Note that the behavior of the $D$ gate is undefined on all the states except
$|0^m\rangle$. In this paper, it is irrelevant what happens to these
states as the corresponding qubits go to a sink gate after the
application of the $D$ operator. 
Also, the behavior of the $D$ gate at 
$|\amp{0^m}|=\delta$ is irrelevant in this paper.

If the $m$ qubits
passed through the $D$ gate are part of a larger quantum system
with $n>m$ qubits, then the
$D$ operator behaves as above for all amplitudes $\amp{j;0^{m}}$,
where $j, 0\leq j \leq 2^{n-m}$. Thus, amplitudes of different states
$|j;0^{m}\rangle$ converge to $0$ or $1$ 
depending on the initial value of $\amp{j;0^m}$, independently of each other. 

In practice, only an approximate behavior of $D$ may be realizable 
within finite convergence time (Figure \ref{fig:Dgate}). This can
be sufficient, depending on the circuit which uses the gate. 
We will see that $\delta_0$, $\delta$, and $\delta_1$
are defined by 
the types of problems we are trying to solve.

The $D$ gate can be intuitively
thought of as a contractive operator that evolves general states towards a
single (stable) state $|0^m\rangle $. Clearly, it cannot be realized in an
isolated system where the only permissible operators are unitary. However,
as discussed before, a truly isolated system is, in any case, a theoretical
ideal which is difficult to realize in practice. 
In Section \ref{section:Dgate}, we see that the $D$ gate needs
inherently nonlinear behavior as well as dissipative behavior.
It is similar to the speculative operator 
defined by Abrams and Lloyd \cite{AL98}; however, the 
nonlinearity depends only on the amplitude and not on phases.

To meet the requirement of having at most a polylogarithmic number of
qubits, we use a dense encoding of $n$ classical bits, labelled
$x_0,\ldots,x_{n-1}$, in only $O(\log n)$ qubits.  There are several
alternatives. The following simple one suffices. 
Assume $n$ is a power of
$2$. Interpret the states of $\log n$ qubits as addresses and if the $j$th
classical bit is $1$, then $\amp{j}=1$, else $\amp{j}=0$.   
As before, the value $1$ represents some appropriate constant $c$;
if the system has $\log n$ qubits, $c=1/\sqrt{n}$ is sufficient.
We include a special \emph{sink state} $z$ so that the
probabilities add to $1$. Thus, 
$|\amp{z}|^2 = 1-\sum_{j=0}^{n-1} |\amp{j}|^2$. 
The sink state may be composed of several states.
For example, if we think of the sink state as an additional qubit,
then $\amp{j;0}$ can be used for encoding $0$ and $1$, and
$\sum_{j=0}^{n-1} |\amp{j;1}|^2 = 1-\sum_{j=0}^{n-1} |\amp{j;0}|^2$. 

As in Deutsch's quantum computational network model \cite{Deu89}, our
model has \emph{quantum gates} interconnected by \emph{wires}.  In
particular, we preserve \emph{source} and \emph{sink} gates%
\footnote{When a qubit goes to a sink gate, it is measured to remove the 
entanglements with the remaining qubits.}.  
Our model also has the standard reversible unitary operator $U$.
Since we are working within the constraints of polylogarithmic
qubits, we allow arbitrary unitary operators\footnote{With polynomial
number of qubits, if we allow arbitrary unitary operators (of size
$2^{n^k} \times 2^{n^k}$), then the cost of constructing the apparatus
can defeat the gains in computational speed.}.
Of course, a unitary operator of dimension
$2^{\log^{k} n}, k\geq 1$, can be approximated by decomposition into
$\log^{k} n, k\geq 1$, local operators using standard techniques \cite{BBC95}.
However, in principle, any
unitary matrix is allowed in quantum mechanics. 
We say that a $U$ gate has \emph{precision} $p$ 
if all the values in the matrix can be approximated to 
within $1/2^{p+1}$ by binary rational numbers.

The unitary operator implicitly
contains the important properties that overcome the main problems
with classical parallel computing: communication bottleneck and 
synchronization overhead. 
Because of entanglement, operations on even one qubit can affect
all the qubits instantaneously,
without the need for explicitly wiring them together. 
While the research in literature has focussed on exploiting the exponential
properties of quantum systems to perform computations
infeasible with the current digital systems,
our model is using only the entanglement and interference present in
quantum systems to overcome the communication bottleneck 
and synchronization problems that
have plagued the implementation of parallel computers.
We believe that scalable parallel computers can be built
using quantum systems.

A \emph{quantum neural network} $\QNN(s(n),d(n))$ of precision $p(n)$
is a circuit of size $s(n)$
and depth $d(n)$, constructed from the gates $D$ and $U$ of 
precision $p(n)$. 
Size denotes the number of qubits in the circuit and depth denotes
the longest sequence of gates from input to output.
In general, the reversible $U$ gate is followed by the irreversible $D$
gate to eliminate the problem of localization. 
Usually, the precision of the circuits will be $O(s(n))$.

Size, depth, and precision are important complexity-theoretic measures
that quantify various aspects of computations. 
Depth corresponds to the number of steps needed to solve a given problem.
Size usually corresponds to the size of apparatus; precision 
also characterizes the apparatus needed for solving the problem.


\subsection{Examples of QNNs}

We give examples of QNNs for some simple circuits with the universal \NAND\
gates. Define a unitary matrix $U_{nand}$ that operates on two
qubits as follows.

\[ U_{nand} = 
\left[
\begin{array}{cccc}
\frac{1}{\sqrt{6}} & 0 & \frac{1}{\sqrt{6}} & \frac{-2}{\sqrt{6}} \\
\frac{1}{\sqrt{3}} & \frac{1}{\sqrt{3}} & -\frac{1}{\sqrt{3}} & 0 \\
\frac{1}{3\sqrt{2}}  & \frac{\sqrt{2}}{3} & \frac{1}{\sqrt{2}} & \frac{\sqrt{2}}{3} \\
\frac{2}{3} & -\frac{2}{3} & 0 & \frac{2}{6} 
\end{array}
\right] 
\begin{array}{r}
|00\rangle \\
|01\rangle \\
|10\rangle \\
|11\rangle 
\end{array}
\hspace*{3em}
X = \frac{1}{2}
\left[
\begin{array}{l}
x_1\\
1\\
x_2\\
1
\end{array}
\right]
\]

\begin{figure}
\begin{center}
\input{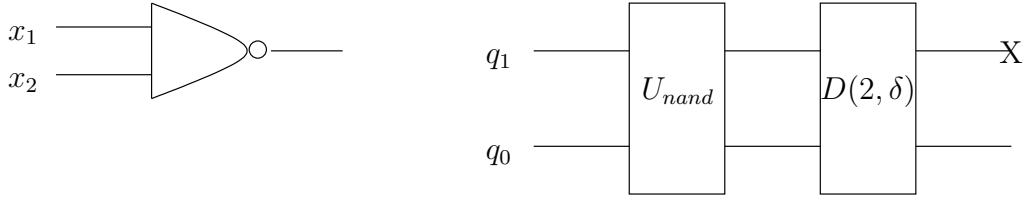}
\caption{QNN for $\NAND(x_1,x_2)$}
\label{fig:NAND}
\end{center}
\end{figure}

In the input vector $X$, $0$ is encoded as $0$ and $1$ is encoded
as $1/2$. 
The amplitude of the state $|00\rangle$ in the state vector
$U_{nand} X$ is $0$ if and only if $\NAND(x_1,x_2)=0$. 
Thus, when we pass the qubits through
a $D(2,\delta)$ gate with $0 < \delta < 1/2\sqrt{6}$, 
the amplitude of state 
$|00\rangle$ has the correct answer (Figure \ref{fig:NAND}).
We have not included the sink state in the above description.

\begin{figure}
\begin{center}
\input{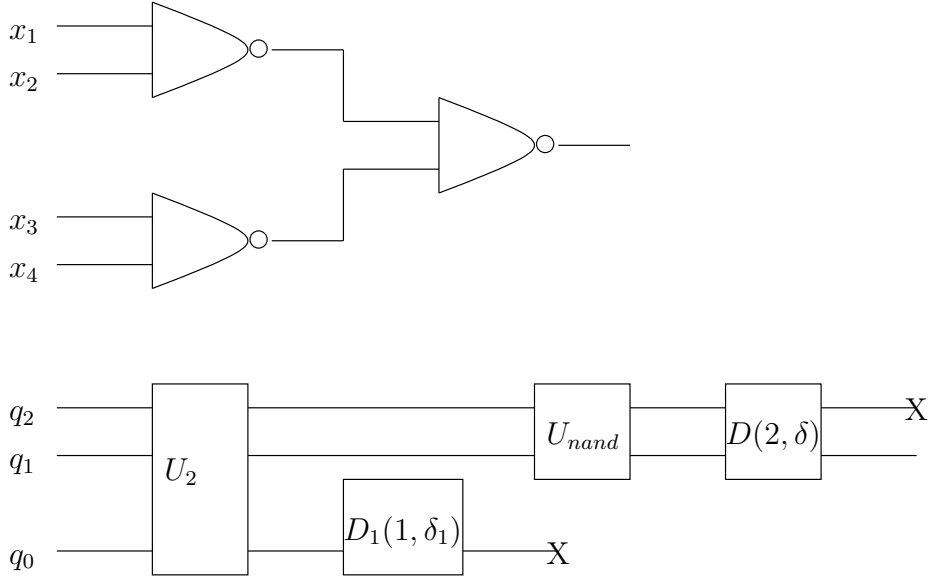}
\caption{QNN for a circuit with NAND gates}
\label{fig:NANDcircuit}
\end{center}
\end{figure}

Next we show how a circuit with three NAND gates can be
simulated by a QNN (see Figure \ref{fig:NANDcircuit}).
Define $U_2$ to be the following unitary matrix.
\[
U_2 = 
\left[
\begin{array}{ll}
U_{nand} & \mathbf{0}\\
\mathbf{0} & U_{nand}
\end{array}
\right]_{8 \times 8}
\]
The operator  $U_2$ operates on 
$X^T = 1/\sqrt{8} [x_1, 1, x_2, 1, x_3, 1, x_4, 1]$. When the result is
passed through the $D_1(1,\delta_1)$ gate, $0 < \delta_1 < 1/\sqrt{6\cdot 8}$,
we simulate the two
NAND gates simultaneously. The resulting vector is
$[\NAND(x_1,x_2),1,\NAND(x_3,x_4),1]$, 
which is used to simulate the last NAND gate as before. 
As an aside, the construction of QNNs for arbitrary circuits of 
NAND gates, given in Section \ref{section:power}, is different 
from the construction given in Figure \ref{fig:NANDcircuit} 
and has worse bounds.

\section{Computational Power of QNNs}
\label{section:power}

In this  section we characterize the computational power of QNNs using
the known results from the complexity theory of parallel computations and 
threshold circuits (used for modeling neural networks \cite{MP43,MP47}).
First, we define a class of circuits called \EC\ and show that
with polynomial size  they have the same computational
power as threshold circuits. \EC\ circuits can be converted into
QNNs using a proper encoding of classical bits into qubits.

\subsection{Circuits}
Let $G$ be a set of gates (functions) that map several bits to one
bit. 
For each $n\geq 0$, a circuit
$C_n$ over the set $G$ is a directed, acyclic graph with a list of input
nodes (with no incoming edges), 
a list of output nodes (with no outgoing edges), 
and a gate in $G$ labeling each
non-input node. Given a binary input string $(x_1,\ldots,x_n)$, we
label each input node $x_i$ or $\bar{x_i}$ (NOT $x_i$). 
For every other
node $v$ with $m$ predecessors $y_1,\ldots,y_m$, we recursively
assign a value $g(y_1,\ldots,y_m)$, where $g$ is the gate that labels
node $v$.  $C_n$ outputs the value given by the list of output nodes.

Unless otherwise stated, in this paper, we shall allow the circuits to
have arbitrary fan-in and fan-out. Circuits with fan-out $1$ will be
called ``opened circuits,'' where the corresponding graph is a tree.
The \emph{size} of a circuit $C_n$ is the number of nodes
in $C_n$ and the \emph{depth} of
$C_n$ is the length of the longest path from any input node to an
output node. 

A \emph{circuit family} is an infinite list of circuits 
$\mathbf{C}=(C_1,C_2,\ldots,C_n,\ldots)$ where $C_n$ has $n$
binary inputs. $\mathbf{C}$ computes a family of boolean functions
$(f_1,f_2,\ldots,f_n,\ldots)$, where $f_n$ is the boolean function
computed by circuit $C_n$. We say that $\mathbf{C}$ has 
\emph{size complexity} $s(n)$ and \emph{depth complexity} $d(n)$
if for all $n\geq 0$ circuit $C_n$ has size at most $s(n)$ and depth
at most $d(n)$. 
Size and depth are important complexity descriptions of circuits that
respectively characterize the size of apparatus and the number of steps
needed to compute a family of boolean functions. 

A circuit family is \emph{polytime-uniform} if there exists a Turing
machine with time bound $n^{k}, k\geq 1$, that constructs circuit 
$C_n$ on input $1^n$. 
A circuit family is \emph{logspace-uniform} if there exists a Turing machine
with space bound $\log n$ that constructs circuit $C_n$ on input $1^n$.
Uniformity conditions capture the complexity of constructing the
circuit to compute boolean functions with inputs of a given size.

\subsection{Threshold Circuits}
A \emph{threshold function} with threshold $\Delta$ 
is a boolean function denoted 
$\th^{n,\Delta} : \{0,1\}^n \rightarrow \{0,1\}$ such that
\[\th^{n,\Delta}(x_1,\ldots,x_n)= \left\{
	\begin{array}{ll}
	1 & \mbox{if } \sum_{i=1}^n x_i \geq \Delta \\
	0 & \mbox{otherwise}
	\end{array} \right.
\]
for $x_1,\ldots, x_n \in \{0,1\}$, where $0\leq \Delta \leq n$.
A \emph{weighted threshold function} of weight bound $w$ and threshold
$\Delta$ is a  boolean function denoted
$\th^{n,\Delta}_{w_1,\ldots,w_n}: \{0,1\}^n \rightarrow \{0,1\}$ such
that
\[\th^{n,\Delta}_{w_1,\ldots,w_n}(x_1,\ldots,x_n)= \left\{
	\begin{array}{ll}
	1 & \mbox{if } \sum_{i=1}^n w_i x_i \geq \Delta \\
	0 & \mbox{otherwise}
	\end{array} \right.
\]
for $w_1,\ldots,w_n, \Delta \in \mathcal{Z}$ (set of integers), $|w_i| \leq w$
for all $i$, and $x_1,\ldots, x_n \in \{0,1\}$. 

A \emph{threshold circuit} is a circuit over the set of
threshold gates. Let $\TC(s(n),d(n))$ denote the collection of threshold
circuits of size $s(n)$ and depth $d(n)$. 
We overload the
notation \TC\ to also denote the class of functions computable by
these circuits. \emph{Weighted} $\TC(s(n),d(n))$ of weight bound $w$
denotes the collection of threshold circuits using weighted threshold
gates of weight bound $w$ of size $s(n)$ and depth $d(n)$.

\bthm\cite{RT92}
Suppose an analytic function $f(x)$ has a convergent Taylor series
expansion $f(x)=\sum_{n=0}^{\infty} c_n (x-x_0)^n$ over an
interval $|x-x_0|\leq \epsilon$ where $0< \epsilon < 1$, and the
coefficients are rationals $c_n=a_n/b_n$ where $a_n, b_n$ are integers
of magnitude at most $2^{n^{O(1)}}$. Then polytime-uniform threshold
circuits of polynomial size and simultaneous constant depth can compute
$f(x)$ over this interval within accuracy $2^{-n^{c}}$ for any constant
$c\geq 1$.
\ethm

The above theorem implies that $\TC(n^{O(1)},O(1))$ can
approximate elementary functions such as sine, cosine, exponential,
logarithm, and square root. $\TC(n^{O(1)},O(1))$ can also exactly
compute integer and polynomial quotient and remainder, interpolation
of rational polynomials, banded matrix inverse, and triangular 
Toeplitz matrix inverse \cite{RT92}.

\subsection{Equality (Threshold) Circuits}
A (weighted) equality threshold function of weight bound $w\geq 0$ is a
boolean function denoted $\eq^n_{w_1,\ldots,w_n} : \{0,1\}^n
\rightarrow
\{0,1\}$ such that
\[\eq^n_{w_1,\ldots,w_n}(x_1,\ldots,x_n)= \left\{
	\begin{array}{ll}
	0 & \mbox{if } \sum_{i=1}^n w_i x_i = 0 \\
	1 & \mbox{otherwise}
	\end{array} \right.
\]
for $w_1,\ldots,w_n\in \mathcal{Z}$, $|w_i| \leq w$ for all $i$,
 and $x_1,\ldots, x_n \in \{0,1\}$.
Note that \eq\ gates possess an elegant
algebraic structure as they can be naturally generalized to any 
domain, such as complex numbers.

An \emph{equality threshold circuit} of weight bound $w$ is a circuit over
the set of equality threshold gates such that all the weights 
are absolutely bounded by $w$. 
Let
$\EC(s(n),d(n))$ denote the collection of equality circuits of size
$s(n)$ and depth $d(n)$. \EC\ will also be used to denote the class of
functions computable by these circuits.

We will see that the weights of a \EC\ circuit can be encoded 
appropriately in unitary matrices, and quantum entanglement and interference 
can be used to compute $\sum_{j=1}^n w_j x_j$ in one step. 
Thus, \EC\ provides a good model to help combine ideas from
quantum computing and threshold circuits.

\subsection{AND-OR Circuits}
\NC\ is the collection of logspace-uniform circuits of size $n^{O(1)}$ and 
depth $(\log n)^{O(1)}$ over the set $\{$AND, OR$\}$ of gates with
fan-in bounded by $2$. It is the class of problems that can
theoretically be solved efficiently in parallel and contains numerous
natural problems.  We write \NC\ to denote both the class of
circuits and functions computable by the circuits.
$\NC^i, i\geq 1$, denotes the subclass of \NC\ where the depth is limited to
$O(\log^i n)$.  Though \NC\ circuits use remarkably fast AND, OR gates,
the implementations of circuits have not been successful because of the
communication bottleneck and synchronization cost involved in wiring
a polynomial number of processors. The algorithms that show that problems
are in \NC\ are typically not used to solve problems in parallel;
new algorithms are designed depending on the parallel architecture
model that takes communication cost into account.

$\NC^1$ is known to have several natural problems such as integer
arithmetic and matrix  multiplication. 
$\NC^2$ includes matrix inverses
and matrix rank computation as well. Please see any standard text such
as \cite{Koz91} and \cite{Rei93} for more details.

\subsection{Computational power of $\TC$ and $\EC$ circuits}
We now show that \TC\ and \EC\ circuits have essentially the same
computational power if we consider circuits of polynomial size.
In all our constructions, if the
original circuits are polytime-uniform and logspace-uniform, the
constructed circuits are also polytime-uniform and logspace-uniform,
respectively.

\blem\ 
\begin{enumerate}
\item $\th^{n,\Delta}(x_1,\ldots,x_n)$ can be simulated by $\EC(O(n),2)$
of weight bound $O(n)$.

\item 
$\TC(s(n),d(n)) \subseteq \EC(s^3(n),d(n)+1)$ of weight bound
$O(s^2(n))$, and 
$\TC(s(n),d(n)) \subseteq \EC(O(s^2(n)),2d(n))$ of weight bound $O(s(n))$.

\end{enumerate}
\elem

\begin{figure}
\begin{center}
\input{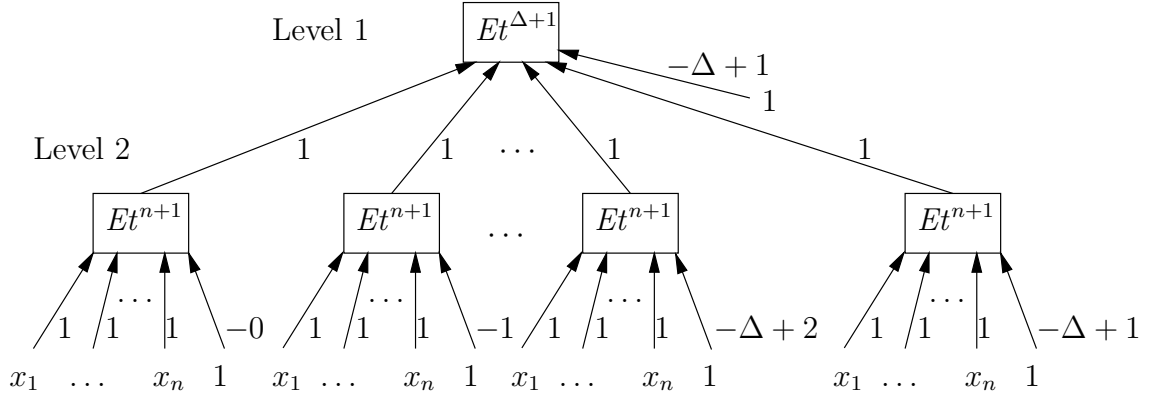}
\caption{Simulation of Threshold gate $\th^{n,\Delta}$ by \EC\ of weight 
bound $O(n)$}
\label{fig:ThEq}
\end{center}
\end{figure}

\begin{figure}
\begin{center}
\input{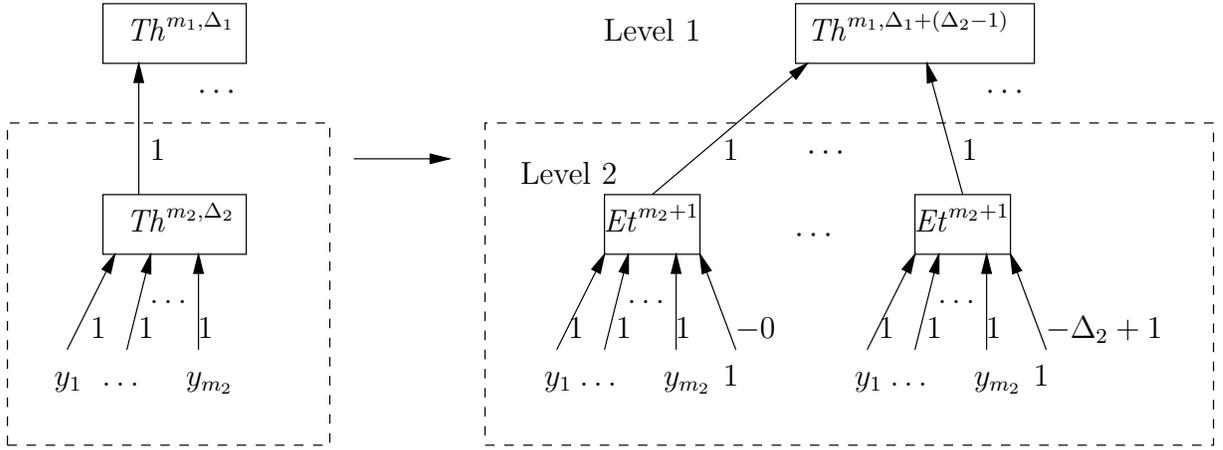}
\caption{Simulation of \TC\ with \EC}
\label{fig:TCEq}
\end{center}
\end{figure}

\noindent\textbf{Proof. \ } 
\begin{enumerate}
\item 
Figure \ref{fig:ThEq} shows the \EC\ circuit that simulates the
threshold gate $\th^{n,\Delta}(x_1,x_2,\ldots,x_n)$.  
Note that the gates at level $2$ check whether $\sum_{i=1}^n x_i$ is one of
the values $0,1,\ldots, \Delta -1$, respectively. 
\begin{description}
\item[Case ($\sum_{i=1}^n x_i < \Delta$):] 
Exactly one gate at level $2$ outputs $0$ and all others output $1$ giving
the cumulative output of $\Delta-1$,
which when combined with the input $-\Delta +1$ at level $1$ gives
the output of $0$. 
\item[Case ($\sum_{i=1}^n x_i \geq \Delta$):] 
All the gates at level $2$ output $1$ giving the cumulative output of 
$\Delta$, which when added
to the $-\Delta+1$ input at level $1$ gives the final output of $1$.
\end{description}
Thus, the cumulative output of gates at level $2$ is 
$\Delta$ and $\Delta -1$ if the threshold gate outputs $1$
and $0$, respectively. The level $1$ gate is used to add the
outputs of gates at level $2$ and subtract
$\Delta -1$ from this sum.

The weight bound of the constructed circuit is $|-\Delta + 1| = O(n)$.

\item 
An obvious replacement of each threshold gate by the circuit of
Figure \ref{fig:ThEq} gives a depth of $2d(n)$ and size $O(s^2(n))$.
We can reduce this 
bound to $d(n)+1$. Consider a threshold gate $\th^{m_2,\Delta_2}$
connected to another threshold gate $\th^{m_1,\Delta_1}$. 
Figure \ref{fig:TCEq} shows how the threshold gate at level $2$
can be replaced by the equality threshold gates at level $2$ described
in Figure \ref{fig:ThEq}. Since the cumulative output of these
equality threshold gates is $\Delta_2$ and $\Delta_2 -1$ instead of
$1$ and $0$, we simply increase the threshold of the level $1$
gate of Figure \ref{fig:TCEq} by $\Delta_2 -1$. 
It is easy to see
that this replacement does not affect the function computed by the
circuit.

The construction is as follows. Starting at the input level 
of \TC\ circuit, replace each threshold gate as described above
and adjust the threshold at the next level appropriately. 
Repeat the process for all levels.
At the output level, use both level $1$ and $2$ gates of Figure
\ref{fig:ThEq} so that the final gate is also a \eq\ gate. 
As shown in Figure \ref{fig:TCEq}, 
connect the new gates to all the gates connected
by the replaced threshold gate. 

Since the maximum threshold for all the threshold gates is bounded
by $s^2(n)$ and each threshold gate is replaced by $\Delta$ \eq\ gates,
the final circuit has $O(s^3(n))$ \eq\ gates.
The weight bound for the circuit is $O(s^2(n))$ because at the output
level, the thresholds of all the $s(n)$ threshold gates are added 
together.
\end{enumerate}
\eproof

\blem\ 
\begin{enumerate}
\item $\eq^n_{w_1,\ldots,w_n}(x_1,\ldots,x_n)$ can be simulated by 
weighted $\TC(3,2)$ of weight bound \\
$\max(|w_1|,\ldots,|w_n|)$.

\item $\EC(s(n),d(n))$ of weight bound $w$ $\subseteq$ Weighted 
\ $\TC(2s(n)+o,d(n)+1)$ of
weight  bound $w$, where $o$ is the number of output nodes in the 
$\EC(s(n),d(n))$ circuit.
\end{enumerate}
\elem

\begin{figure}
\begin{center}
\input{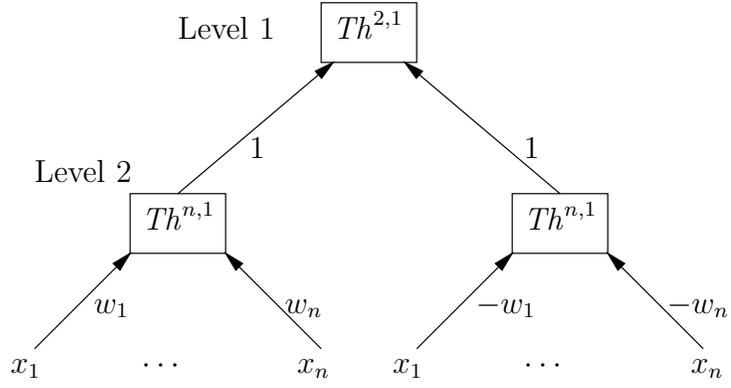}
\caption{Simulation of Equality threshold gate $\eq^n$ by weighted $\TC(3,2)$}
\label{fig:Eq2Th}
\end{center}
\end{figure}

\begin{figure}
\begin{center}
\input{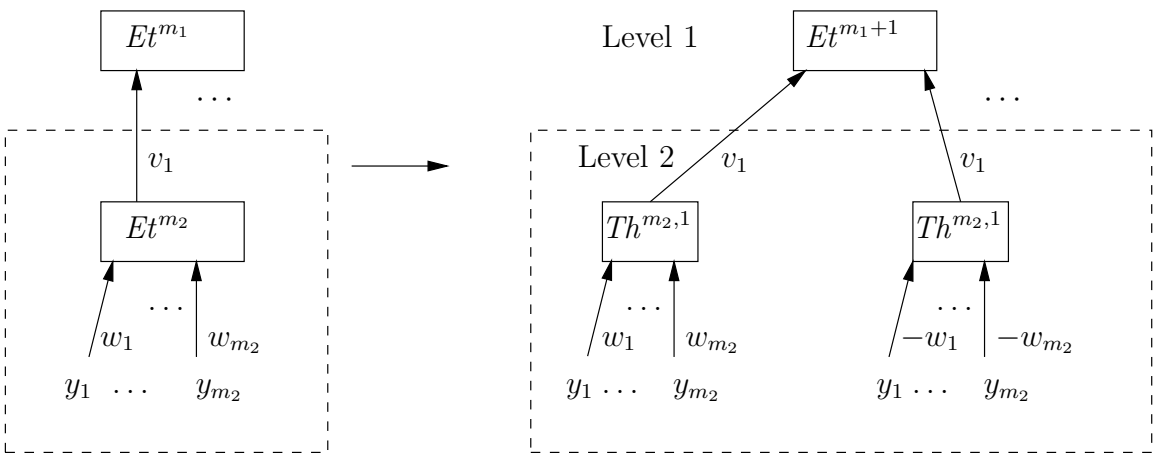}
\caption{Simulation of \EC\ with \TC}
\label{fig:EC2TC}
\end{center}
\end{figure}

\noindent\textbf{Proof. \ } 
\begin{enumerate}
\item 
Figure \ref{fig:Eq2Th} shows the $\TC(3,2)$ circuit that simulates
the gate
$\eq^n_{w_1,\ldots,w_n}(x_1,\ldots,x_n)$.
The leftmost gate at level $2$ outputs $1$ if
$\sum_{i=1}^n w_i x_i \geq 1$ and the rightmost gate
outputs $1$ if $\sum_{i=1}^n w_i x_i \leq -1$.
The gate at level $1$ adds the two outputs from level $1$
gates and outputs $0$ if and only if $\sum_{i=1}^n w_i x_i = 0$.

\item 
An obvious replacement using Figure \ref{fig:Eq2Th} gives a circuit
of  depth $2d(n)$. 
We can reduce this bound to $d(n)+1$. 
Consider an equality threshold gate 
$\eq^{m_2}$ connected to another equality threshold gate $\eq^{m_1}$. 
Figure \ref{fig:EC2TC} shows how the equality threshold gate at level $2$
can be replaced by the equality threshold gates at level $2$ described
in Figure \ref{fig:Eq2Th}.

The construction is as follows. Starting at the input level 
of \EC\ circuit, replace each $\eq$ gate with the level $2$
threshold gates given in Figure \ref{fig:Eq2Th}.
Repeat the process until the output level. 
At the output level, use both level $1$ and $2$ gates of Figure
\ref{fig:Eq2Th} so that the final gate is also a \th\ gate. 
As shown in Figure \ref{fig:EC2TC}, 
connect the new gates to all the gates connected
by the replaced threshold gate. 

Each \eq\ gate is replaced by two \th\ gates except at the output level
where each \eq\ gate is replaced by $3$ \th\ gates.

Since we do not change the absolute value of the
weights, therefore the weight bound of
the new circuit is the same as that of the original circuit.
\end{enumerate}
\eproof

\blem
\item Weighted $\TC(s(n),d(n))$ of weight bound $w$ 
$\subseteq \TC(O(w \cdot s(n)),d(n))$.
\elem

\noindent\textbf{Proof. \ } 
\begin{enumerate}
\item 
Suppose a threshold gate $A$ is connected to another gate $B$ with
an edge with weight $w_{AB}\geq 0$. Simply make $w_{AB}$
copies of gate $A$ and connect them to gate $B$ using an edge of 
weight $1$. In general, we need at most $w \cdot s(n)$ copies.

Edges of weight $0$ can be removed from the circuit.
If a particular edge weight is negative, we can make it
positive using the following standard technique. 

Repeat the following steps for all gates at all levels starting from the
output level down to the input level.

\begin{enumerate}
\item 
If any weight $w_i<0$ is negative, replace the weight
with $-w_i >0$ and the corresponding input $x_i$ with $1-x_i$
as shown in Figure \ref{fig:negative1}. Decrease the threshold of
the gate by $w_1$. Thus, $\sum_{i=1}^m w_i x_i \geq \Delta$ if and
only if $-w_1(1-x_1) + \sum_{i=2}^m w_i x_i \geq \Delta-w_1$. 

\begin{figure}
\begin{center}
\input{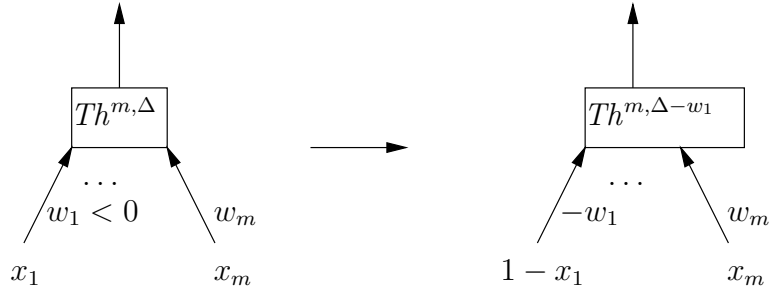}
\caption{Making weights positive by complementing input}
\label{fig:negative1}
\end{center}
\end{figure}

With this step, some gates at intermediate levels may have two
labels $x_i$ and $1-x_i$. If this is the case, duplicate the gate
labelling one $x_i$ and the other $1-x_i$ retaining all the input
connections. In the worst case, we double the number of gates.

\item
To complement the output of a gate from $y$ to $1-y$, use the
process shown in Figure \ref{fig:negative2}. Replace all the weights
$w_i$ with $-w_i$ and change the threshold to
$1-\Delta$. Thus, the gate outputs $1-y$ if and only if 
$\sum_{i=1}^m w_i x_i < \Delta$ if and only if 
$-\sum_{i=1}^m w_i x_i > -\Delta$ if and only if 
$\sum_{i=1}^m (-w_i) x_i \geq 1-\Delta$.
\end{enumerate}

\begin{figure}
\begin{center}
\input{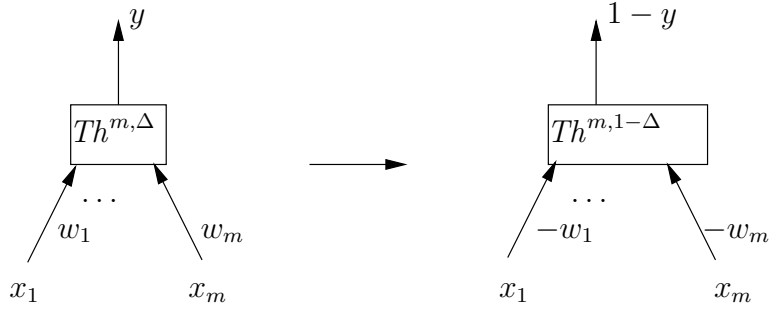}
\caption{Complementing output by negating weights}
\label{fig:negative2}
\end{center}
\end{figure}

When the input level is reached, change $x_i$ to $\bar{x_i}$
and vice versa as needed. 
The above process does not change the weight bound of the circuit.
\end{enumerate}
\eproof

\blem\ 
\begin{enumerate}
\item \NAND$(x_1,x_2)$ can be simulated by $\eq^3_{1,1,-2}(x_1,x_2,1)$.

\item For all $i\geq 1$, $\NC^i\ \subseteq \EC(n^{O(1)}, O(\log^i n))$ of
weight bound $2$.

\item \cite{Ruz81} $\mathrm{SPACE}(s(n)) \subseteq $ {\rm AND-OR} 
circuits of size
$2^{s(n)}$ and depth $s^2(n)$ via $s^2(n)$-uniform space-bounded 
Turing machines.

\end{enumerate}
\elem

\noindent\textbf{Proof. \ } 
\begin{enumerate}
\item Follows immediately from the definitions.

\item Since NAND gate is a universal gate, it can be used in place of
AND and OR gates for \NC\ circuits. Then, use the 
$\eq^3_{1,1,-2}(x_1,x_2,1)$ gate to simulate NAND gates. 
\end{enumerate}
\eproof

We are now ready to characterize the computational power of \QNNs.
Recall that the
precision denotes the number of bits used for characterizing the
amplitudes. Thus, if an amplitude is defined with precision $p$,
then the actual value may differ by $1/2^{p+1}$, that is, the errors must be
less than $1/2^{p+1}$.

\begin{figure}
\begin{center}
\input{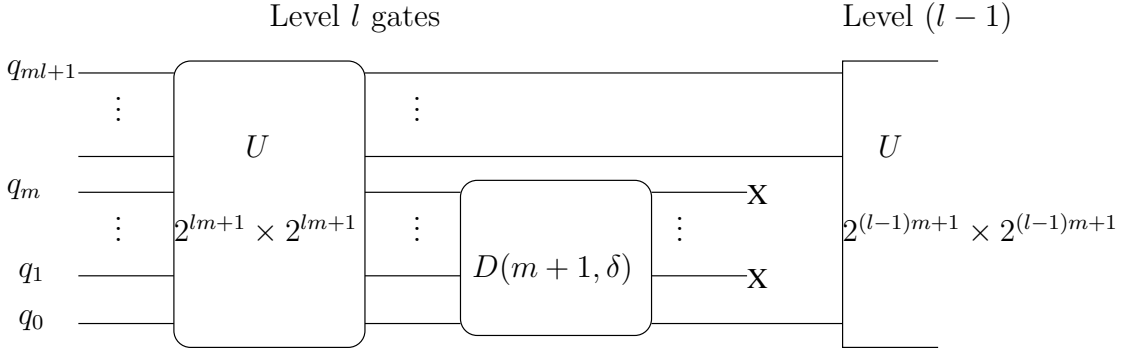}
\caption{QNN for level $l$ of $\EC$ circuit}
\label{fig:onelevel}
\end{center}
\end{figure}

\bthm
$\EC(s(n), d(n))$ of weight bound $w$ 
$\subseteq \QNN(O(d(n) \log s(n)), 2d(n))$ of 
precision $O(\log w + d(n) \log s(n))$.
\ethm
\bproof
For the sake of clarity, we write $s$ and $d$ instead of $s(n)$ and 
$d(n)$, respectively. 
Assume that all the gates have $s=2^m$ inputs each;  add inputs
with weight $0$, if necessary.
Without loss of generality, we may assume that the circuits have one output.
(Otherwise, connect all the output gates to another threshold gate
and remove the corresponding layer in the constructed QNN.)

Given an unbounded fan-in, unbounded fan-out $\EC(s,d)$ circuit 
of weight $w$, starting from the output gate ``open''
the circuit by duplicating gates of fan-out greater than $1$
to obtain a new circuit.
Add dummy gates to obtain a levelled circuit $\EC(O(2^{md}),d)$. 

Given this structured circuit, we will replace all the \EC\ gates
at level $l$ by a $U$ operator of size $2^{lm+1}\times 2^{lm+1}$
followed by a $D(m+1,\delta)$ operator (Figure \ref{fig:onelevel}), 
where the threshold $\delta$ is specified below. 
Thus, we need a system with $md+1 = O(d \log s)$
qubits $q_0,q_1,\ldots, q_{md}, q_{md+1}$. 
After each $D(m+1,\delta)$ operator, 
$m$ qubits (except $q_0$) go to a sink gate so that we have
only one qubit left after $d$ levels of alternating $U$ and $D$ gates. 
Then this qubit is observed along state $|0\rangle$ to obtain
the final answer.

For each level $l$ of the $\EC$ circuit do the following.
\begin{enumerate}
\item
Normalize the weights $w_{jk}, 1\leq k \leq s$, at each gate $j$ 
by dividing them by $\sqrt{\sum_{k=1}^s |w_{jk}|^2}$. 
Since weights are bounded by $w$, therefore 
$\sqrt{\sum_{k=1}^s |w_{jk}|^2} \leq \sqrt{s} w$ and
we require at least $O(\log s + \log w)$ precision for the weights.
(To avoid the accumulation of errors as the computation proceeds
through different levels, we will need slightly higher precision.)

\item Construct a banded matrix $U$ of size $2^{lm+1}\times 2^{lm+1}$ 
and band width $2^{m+1}$.
For each gate $j, 1\leq j \leq 2^{m(l-1)}$,
with normalized weights $w_{j1}, w_{j2}\ldots, w_{js}$, 
we have a block $B_j$ in $U$.

The first row of $B_j$ corresponds to the weights of gate $j$ and the
remaining rows are chosen so that the rows of $B_j$ form a set of 
orthonormal vectors.
The $U$ matrix is the banded matrix with diagonal blocks 
$B_1,\ldots,B_{2^{(l-1)m}}$.
Note that the first row of block $B_j$ computes the weighted sum
$\sum_{k=1}^s w_{jk}x_{jk}$ for the $j$th gate if the input is
$X^T = \left[x_{j1}, 0, x_{j2}, 0, \cdots, x_{j 2^m}, 0\right],
1\leq j \leq 2^{m(l-1)}$.
If we pass the $m+1 =\log s  +1$ least significant qubits 
through a $D(m+1,\delta)$ gate, we simulate all the $\eq$ gates at one
level in parallel.

\[B_j = \left[ \begin{array}{lllllll}
w_{j1}   & 0        & w_{j2}   & 0      & \ldots & w_{j2^m} & 0\\
w_{j_21} & w_{j_22} & w_{j_23} &w_{j_24}& \ldots & w_{j_2 2^{m+1}-1}
& w_{j_2 2^{m+1}} \\
& & & \ldots & & & \\
w_{j_{2^{m+1}} 1} & w_{j_{2^{m+1}} 2} & w_{j_{2^{m+1}} 3} & w_{j_{2^{m+1}} 4}
&  \ldots & w_{j_{2^{m+1}}2^{m+1}-1} & w_{j_{2^{m+1}}2^{m+1}}
\end{array}\right]_{2^{m+1} \times 2^{m+1}} \hspace*{2em}
\]
\[
U = \left[ \begin{array}{llll}
B_1 & & & \\
    & B_2 & & \\
& & \ldots &  \\
& & & B_{2^{(l-1)m}}
\end{array}\right]_{2^{ml+1} \times 2^{ml+1}}
\]

\end{enumerate}

Once the circuit is constructed as above, the computation is performed
as follows.

\paragraph{Computation:}

The input at level $l$ is a state vector of size $2^{ml+1}\times 1$
as follows: $X^T = \left[x_{j1}, 0, x_{j2}, 0, \cdots, x_{j 2^m}, 0\right],
1\leq j \leq 2^{m(l-1)}$, where $x_{jk}, 1\leq k\leq 2^m$, are inputs
to the $j$th gate. 
The operation $U X$ computes the the weighted sum
of all the gates at level $l$.
Pass the $m+1$ least significant qubits through
a $D(m+1,\delta)$ gate followed by the sink gate for $m$
qubits (except $q_0$), where $\delta$ is chosen to be sufficiently small.
(The exact value is specified below and depends on error bounds.)
Note that this process simulates all the $\eq$ gates at one
level in parallel, where $U$ gate computes the weighted sum
and the $D$ gate acts as a zero checker.

\paragraph{Precision of amplitudes in the state vector $x$:}

At the lowest level of the circuit we have $2^{md}$ input nodes 
such that $\sum_{i=1}^{2^{md}} |x_i|^2 =1$.
Thus, we need a precision of at least $O(dm)= O(d \log s)$ 
for the state vector. Of course, after every $D$ gate, $1/2^m$
of these states disappear. Therefore, if the amplitudes are not
lost in the sink state, we need a precision of $O(l \log s)$ at
level $l$.

\paragraph{Precision of weights in $U$ gates:}

Let $\epsilon_l\geq 0$ denote the error bound
in weight values at level $l$, where the output node
corresponds to level $1$. 
As mentioned before,
we need to ensure that $\epsilon_l \leq \frac{1}{\sqrt{s}w}$ at
all levels $l$.
Let $w_1,w_2,\ldots,w_s$ be the correct
normalized weights corresponding to a row of $U$. 
Let $w_1^{\prime},w_2^{\prime},\ldots,w_s^{\prime}$ be the erroneous
weights. Let $x_j$ and $x_j^{\prime}, 1\leq j\leq s$, respectively
denote the correct 
and the erroneous input values corresponding to the row of $U$.
Then,
\begin{eqnarray*}
\epsilon_l & \leq & \left |\sum_{j=1}^s w_j^{\prime} x_j^{\prime}
			- \sum_{j=1}^s w_j x_j \right| \\
& \leq & \sum_{j=1}^s(|w_j|+\epsilon_{l+1})(|x_j|+\epsilon_{l+1})
			- \sum_{j=1}^s |w_j| |x_j|\\
& \leq & s \epsilon_{l+1}^2 + \epsilon_{l+1} \sum_{j=1}^s (|w_j|+|x_j|) \\
& \leq & s \epsilon_{l+1}^2 + 2 \epsilon_{l+1} \sqrt{s},
  \mbox{  as $\sum_{j=1}^s |w_j|^2 = 1$ and $\sum_{j=1}^s |x_j|^2 \leq 1$}\\
& \leq & (\sqrt{s} \epsilon_{l+1})^2 + 2 \sqrt{s} \epsilon_{l+1}
\end{eqnarray*}
If $\epsilon_{l} \leq \frac{1}{3^l s^{l/2} w}$, then we have
$\epsilon_l \leq \frac{1}{\sqrt{s} w}$ for all $l$.
Thus, the circuit needs $O(l \log s + \log w)$ precision 
for the weights in $U$ gate at level $l$.

\paragraph{Threshold and Precision for the $D$ gate:}
The smallest integral value that can put the corresponding \eq\ gate
over the threshold is $1$. 
The corresponding weights in \QNN\ are normalized
by dividing by $\sqrt{\sum_{k=1}^s |w_{jk}|^2} \leq \sqrt{s}w$.
Also the normalized input of $1$'s are encoded as
$\sqrt{1/2^{ml}} = 1/\sqrt{s^l}$. 
Therefore, we can choose 
$0< \delta <  1/ \sqrt{s^l}\sqrt{s}w$, a good value being 
$\delta = 1/ 2\sqrt{s^l}\sqrt{s}w$.
With error bound $\epsilon$
we can ensure that the amplitude values never fall in the range
$(\delta_0, \delta_1)$, where $\delta_0 = \epsilon$
and $\delta_1= 2\delta - \epsilon$. 

The output of $D$ gate must be within $\epsilon_l$ of the correct
value at level $l$.
\eproof

\bcol
$\TC(n^{O(1)},O(1))$ $\subseteq$ $\QNN(O(\log n), O(1))$ of 
precision $O(\log n)$.
\ecol

Thus, $\QNN(O(\log n), O(1))$ can
approximate elementary functions such as sine, cosine, exponential,
logarithm, and square root. It can also exactly
compute integer and polynomial quotient and remainder, interpolation
of rational polynomials, banded matrix inverse, and triangular 
Toeplitz matrix inverse.

\bcol
$\NC^1$ $\subseteq$ $\QNN(O(\log^2 n), O(\log n))$ of 
precision $O(\log^2 n)$.
\ecol

\bcol
$\NC^2$ $\subseteq$ $\QNN(O(\log^2 n), O(\log^2 n))$ of 
precision $O(\log^2 n)$.
\ecol

Thus, $\QNN(O(\log^2 n), O(\log^2 n))$ can compute various 
matrix operations such as inverse and rank.

\bcol
$\NC$ $\subseteq$ $\QNN(\log^{O(1)} n, \log^{O(1)} n)$ of 
precision $\log^{O(1)} n$.
\ecol

\bcol
{\rm POLYLOGSPACE} $\subseteq$ $\QNN(\log^{O(1)} n, \log^{O(1)} n)$ of 
precision $\log^{O(1)} n$.
\ecol

Hence, the several natural problems in $\TC$ and $\NC$ 
can be solved using only $\log^{O(1)} n$ qubits
and at most $\log^{O(1)} n$ steps.

So far the $D$ gate was not completely defined. However, to delimit
the power of QNNs, we need to define the behavior completely. For the
following results, we assume that only the least significant qubit out
of the $D$ gate is not connected to the sink gate and
$\amp{j;1} = 0$ for all $j$. 
These assumptions ensure that the $D$ gate is not performing
difficult computations on the undefined states. 
The result holds if the 
behavior of the $D$ gate on all the states $|i\rangle, 1\leq i\leq 2^m-1$,
is the same as that on the state $|0^m\rangle$.

\bthm
$\QNN(s,d)$ of precision $p$ $\subseteq$ 
$\EC(\lceil d/2 \rceil 2^s,\lceil d/2\rceil)$ of weight 
$O(2^{2\lceil\log s +p\rceil})$.
(Read $s(n)$ and $d(n)$ for $s$ and $d$, respectively.)
\ethm
\bproof
Replace every $U$ gate that has $k$ qubits as input with $2^k$ weighted
linear adders and every $D$ gate with a zero checker. Consecutive
linear adders can be combined into one adder, and an adder followed by
the zero checker can be combined into a $\eq$ gate. Thus, the depth
of the $\EC$ circuit is reduced by at least half.

Since the original circuit has $s$ qubits and depth $d$, we need
at most $2^s$ equality threshold gates at each level for a total
of at most $\lceil d/2 \rceil 2^s$ gates. (Multiple $U$ gates at
the same level operating on different qubits need less than 
$2^s$ gates.)

In general, the weights in the $U$ matrix can be complex numbers
while $\eq$ gates are allowed to have only integer weights. Thus,
we need to ensure that both  the real and imaginary parts are zero.
This can be done simultaneously by scaling the imaginary part of weights
by $2^{2\lceil\log s +p\rceil}$ and the real weights by 
$2^{\lceil\log s +p\rceil}$ and then adding them to obtain
a single integer. Scaling by $2^{2{\lceil\log s\rceil}}$ ensures
that the weighted sums of the real and imaginary parts do not mix.
\eproof

Thus, the computational power of simplified
$\QNN(O(\log n), O(1))$ of precision $O(\log n)$ is the same
as that of $\EC(n^{O(1)},O(1))$ of weight bound $n^{O(1)}$ and
$\TC(n^{O(1)},O(1))$.

\subsection{Encoders and Decoders}
In the previous section, we proved that QNNs can process information
in parallel if the information is presented in encoded form. However,
in general, the information may be available only in the classical 
form. In this case, a special apparatus needs to be built to encode
$n$ classical bits into $O(\log n)$ qubits. 
In this section, we show that such an apparatus is indeed feasible
by demonstrating unitary matrices for encoders. These
unitary matrices are highly structured, and hence, in practice their
implementations may look quite different from the way it is presented here.

\paragraph{Encoders:}

Recall that we need to encode $n=2^m$ classical bits $a_i, 0\leq i \leq n-1$,
in $\log n$ qubits as $c \sum_{i=0}^{n-1} \alpha_i |i\rangle$,
where $\alpha_i = a_i$, $a_i \in \{0,1\}$, and $c$ is
an appropriate constant such as $1/\sqrt{n}$.

The unitary operator for the encoders operates on $n$ qubits each in
one of the states $|0\rangle$ or $|1\rangle$, $\log n$ qubits to
be used for encoding, and an additional qubit corresponding to the
environment or sink.  
The operator can be described as
$2^{n+\log n+1} \times 2^{n+\log n+1}$ banded matrix of band size 
$2n \times 2n$.

\[B_{b_{n-1}\ldots b_1 b_0} = c \left[ \begin{array}{llll}
b_0 & e_{1,2} & \ldots & e_{1,2n} \\
1-b_0   & e_{2,2} & \ldots & e_{2,2n} \\
b_1 & e_{3,2} & \ldots & e_{3,2n} \\
& & \ldots &  \\
b_{n-1} & e_{2n-1,2} & \ldots & e_{2n-1,2n} \\
1-b_{n-1} & e_{2n,2} & \ldots & e_{2n,2n} \\
\end{array}\right]_{2n \times 2n} 
\]

\[
U_E = \left[ \begin{array}{llll}
B_{00\ldots 00} & & & \\
& B_{00\ldots 01}  & & \\
&  & \vdots   & \\
& & B_{11\ldots 10} & \\
& & & B_{11\ldots 11}
\end{array}\right]_{2^{n+\log n+1} \times 2^{n+\log n+1}}
\]

All the entries $e_{jk}$ of 
$B_{b_{n-1}b_{n-2}\ldots b_0}$ are chosen so that the rows form
a system of orthonormal vectors.
Now suppose we want to encode $n$ classical bits $a_0,a_1,\ldots,a_{n-1}$
into $\log n$ qubits as described above. Prepare a system in the 
following state $|a_0,a_1,\ldots,a_{n-1}; 0^{\log n} ; 0\rangle$,
where $0^{\log n}$ denotes the state of 
qubits used for encoding and the last qubit
corresponds to the sink state. 
After the operator $U_E$ is applied to this state, we
obtain the following superposition of states:
$c \sum_{i=0}^{n-1} \alpha_i |a_0,a_1,\ldots,a_{n-1}; i; z\rangle$,
where $\alpha_i = a_i$ if $z=0$, and $\alpha_i=1-a_i$ if $z=1$.
Now send the first $n$ qubits corresponding to the classical bits
to sink gates 
to obtain the encoding in the $\log n$ qubits as needed.

\paragraph{Decoders:}
If the output consists of one classical bit, a simple observation
of the corresponding qubit will yield the classical information. Even
if there are more than $1$ bit in the output, they can be observed
after repeated runs of the circuit, or by having multiple copies of
the circuit. Having multiple copies of the circuit has the advantage
of making the computational system robust.


\section{Dissipation Gate}
\label{section:Dgate}

In this section, we show that the behavior of the $D$ gate can be
modeled by cubic nonlinear differential equation. Similar equations
are frequently used in quantum systems of interacting particles. 
Though these equations are not
identical to ours, we believe that they demonstrate the consistency of $D$
gate with the laws of quantum mechanics and identify some possible
approaches for implementing it. Another significant aspect of the $D$ gate
is irreversibility. To achieve this \emph{dissipative} property in a quantum
system, it is necessary to introduce some kind of controlled interaction
with, say, an environment \cite{CL83}. Designing specific mechanisms, though
possible in principle, will not be facile.

\subsection{Amplitude evolution equation}

We show that the dynamical behavior of the $D$ gate on the amplitude of
state $|0^m\rangle $ can be achieved by cubic nonlinearities using stable
points $0$ and $1$. Since it is irrelevant what happens to states
other than $|0^m\rangle$, for simplicity, let us denote $%
\mbox{${\cal
A}(|0^m\rangle)$}$ by $\mathcal{A}$. 
(Indeed, the amplitudes of all the states can evolve as follows without
affecting our results.)
The differential equation that
describes the required behavior is

\begin{equation}
\frac d{dt}\mathcal{A}=R\mathcal{A}\left( \left| \mathcal{A}\right| -\delta
\right) \left( 1-\left| \mathcal{A}\right| \right) \;,  \label{eqn:evol}
\end{equation}
where $R$ denotes the rate of convergence. It is clear that the phase of $%
\mathcal{A}$ remains invariant under this evolution and only its
amplitude is a dynamical variable. So, let us focus on $a\equiv |\mathcal{A}%
| $. Note that $\frac{da}{dt}$ is negative in the range $(0,\delta )$ and
positive in the range $(\delta ,1)$. Thus, $a$ monotonically decreases in
the range $(0,\delta )$ and monotonically increases in the range $(\delta
,1) $. Also, the derivative $\frac{da}{dt}$ is $0$ at points $0,\delta ,1$,
and therefore, $a$ does not change at these points.

Let $a_0$ denote the initial value of $a$. Then the explicit solution to
Equation (\ref{eqn:evol}) is 
\begin{equation}
\left( \frac a{a_0}\right) ^{1/\delta }\left( \frac{a-\delta }{a_0-\delta }%
\right) ^{-1/\delta \left( 1-\delta \right) }\left( \frac{1-a}{1-a_0}\right)
^{1/\left( 1-\delta \right) }=e^{-Rt}  \label{eqn:sol}
\end{equation}
Alternative forms, convenient for analyzing the behavior near the points $%
0,\delta ,1$, may be obtained by raising both sides to the appropriate
power. For example, near $0$, we may write

\begin{equation}
\left( \frac a{a_0}\right) \left( \frac{a-\delta }{a_0-\delta }\right)
^{-1/\left( 1-\delta \right) }\left( \frac{1-a}{1-a_0}\right) ^{\delta
/\left( 1-\delta \right) }=e^{-\delta Rt}  \label{eqn:sol2}
\end{equation}
for convenience.

If $a_0\neq \delta$ and $t\rightarrow \infty $, the right side of the
equation tends to $0$. Since $a$ monotonically decreases in the range $%
(0,\delta)$ and monotonically increases in the range $(\delta,1)$, it
converges to $0$ or $1$ as $t\rightarrow \infty $, depending on whether it
started below or above $\delta$, respectively.

\subsection{Convergence time}

Suppose the system is set up so that amplitudes never fall in the range $%
(\delta_0, \delta_1)$, where $0\leq \delta_0 < \delta < \delta_1 \leq 1$.
The specific values $\delta_0$ and $\delta_1$ depend on the properties of
quantum neural network (QNN) which, in turn, depends on the problems we are
trying to solve. 


The next question is what should be the rate of convergence, $R$, so that
the fixed points $0$ and $1$ are reached with tolerance $\epsilon $ within
some finite time $T$. Demanding both 
\begin{eqnarray}
a(T) &<&\epsilon \quad \text{if}\quad a(0)<\delta _0  \label{tolerance} \\
1-a(T) &<&\epsilon \quad \text{if}\quad a(0)>\delta _1\;,  \label{tol}
\end{eqnarray}
we obtain two equations by substituting $a=\epsilon ,s=\delta _0$ and $%
a=1-\epsilon ,s=\delta _1$ in Equation (\ref{eqn:sol}). The two equations
provide two rates $R_0$ and $R_1$, for $a<\delta _0$ and $a>\delta _1$,
respectively. For our purposes, it is sufficient to choose $R=\max (R_0,R_1)$
so that both inequalities (\ref{tolerance},\ref{tol}) are satisfied. Thus,
given $\delta _0,\delta _1$, $\delta $, $\epsilon $ and time $T$, we obtain
the parameters of the system needed to implement the $D$ gate.

The evolution described so far converges $|\mbox{${\cal
A}(|0^m\rangle)$}|$ to $0$ or $1$ without changing the phase of the state.
In the next step we collapse $\mbox{${\cal A}(|0^m\rangle)$}$ to $0$ or $1$.

\subsection{Dissipation}

Once the amplitude $|\mbox{${\cal A}(|0^m\rangle)$}|$ converges close to $%
0$ or $1$, the quantum system is measured along state $|0^m\rangle $. The
measurement collapses the system to state $|0^m\rangle $ with a high
probability if $|\mbox{${\cal A}(|0^m\rangle)$}|^2$ is close to $1$. Note
that this part of the $D$ operator is also dissipative and irreversible. We
envisage interactions of our quantum system with an environment, in a
controlled fashion, can produce both the nonlinearities and the collapse
needed to construct the $D$ gate.

What happens if the system is observed along state $|0^m\rangle$ and $|%
\mbox{${\cal A}(|0^m\rangle)$}|^2=0$ ? This situation occurs when say a
photon in state $|\rightarrow\rangle$ is observed using an orthogonal filter 
$|\uparrow\rangle$ --- nothing is observed. This difficulty is easily
addressed while solving another problem described below.

In general, the $m$ qubits passed through the $D$ gate are part of a larger
computing system of $m \leq n$ qubits. In this case, we demand the above
nonlinear behavior of the $D$ gate on all $|j;0^m\rangle$ states, for $0\leq
j \leq 2^{n-m}-1$. This general behavior presents the following difficulty.
It is possible that $\sum_{j=0}^{2^{n-m}-1} |\mbox{${\cal A}(|j;0^m\rangle)$}%
|^2 < \sum_{j=0}^{2^{n-m}-1} \sum_{k=1}^{2^m} |\mbox{${\cal A}(|j;k\rangle)$}%
|^2$, and hence, when the system is measured along state $|0^m\rangle$, it
does not collapse to state $|0^m\rangle$ as needed. 

To solve the above problem, we introduce an ancilla qubit $z$ in state $%
|0\rangle$ so that $\mbox{${\cal A}(|j;k;1\rangle)$} =0$ for $0\leq j \leq
2^{n-m}-1$ and $0\leq k \leq 2^m-1$. Once the amplitudes $%
\mbox{${\cal
A}(|j;0^m;0\rangle)$}$ have converged close to $0$ or $1$, transfer the
amplitude of state $|j;0^{m-1}1;0\rangle$ to state $|j;0^m;1\rangle$ for all 
$0\leq j \leq 2^{n-m}-1$. This operation on the last of the $m$ qubits and
the qubit $z$ is performed using the following $4\times 4$ unitary operator.

\[
U = \left[ 
\begin{array}{llll}
1 & 0 & 0 & 0 \\ 
0 & 0 & 1 & 0 \\ 
0 & 1 & 0 & 0 \\ 
0 & 0 & 0 & 1
\end{array}
\right] 
\begin{array}{r}
|00\rangle \\ 
|01\rangle \\ 
|10\rangle \\ 
|11\rangle
\end{array}
\]

The operator $U$ changes the state for the last qubit and
the ancilla qubit from $[b_0,0,b_2,0]$ to $[b_0,b_2,0,0]$ so that when the
last qubit is measured along $|0\rangle$, it collapses to state $|0\rangle$
as $\mbox{${\cal A}(|j;0^{m-1}1;z\rangle)$} =0$ for all $0\leq j \leq
2^{n-m}-1$ and $0 \leq z \leq 1$. Of course, the above unitary behavior and
interaction with an environment can be built in the $D$ gate without
requiring a separate operator.

If we need to collapse all the $m$ qubits to the state $|0^m\rangle$, the
above process can be executed for all the $m$ qubits. However, to solve the
many problems described in this paper, we do not need to do so. All except
the last of the $m$ qubits are discarded to solve these problems. Thus,
it is irrelevant what happens to the other $m-1$ qubits when we observe them.

Next we describe some simple quantum systems that can be modeled using cubic
nonlinearities. While the equations for these systems are different from
Equation (\ref{eqn:evol}), they demonstrate that cubic nonlinearities are
common in modeling quantum systems and indicate some possible approaches for
implementing the $D$ gate. Furthermore, these models clearly demonstrate
that we have not tapped the full power of quantum systems available to us.

\section{Nonlinearities in Quantum Mechanical Systems} 
\label{section:nonlinear}
In this section, we address the question of nonlinearities in quantum
mechanics. While it is generally believed that, at the fundamental level,
the evolution of the entire universe is governed by a linear
Schr\"{o}dinger's equation for ``the wavefunction of the universe,'' this
view is not particularly helpful in practice. For limited physical systems,
such as those found in our laboratories, the linear Schr\"{o}dinger's
equation is only an approximation where certain degrees of freedom can be
ignored or controlled. Thus, in the most celebrated example, an electron in
a hydrogen atom, we find \cite{Sha94}

\begin{equation}
i\hbar \frac{\partial \psi (\mathbf{x},t)}{\partial t}=\left[ -\frac{\hbar ^2%
}{2m}\nabla ^2-\frac{e^2}r\right] \psi (\mathbf{x},t)\;,  \label{Hatom}
\end{equation}
where $2\pi \hbar $ is the Planck's constant, $\psi (\mathbf{x},t)$ is the
wave function (probability amplitude) associated with the electron at
space-time point $(\mathbf{x},t)$, $m$ is the electron mass, $e$ the
elementary charge, and $r$ is the magnitude $\left| \mathbf{x}\right| $.
Here, the nucleus (proton) has been placed at the origin of the coordinate
system, providing only an ``external,'' Coulomb potential ($e^2/r$) in which
the electron moves. All the degrees of freedom associated with the proton
are ignored in this equation. In fact, it is treated as a \emph{%
classical }particle. Had we taken into account its quantum mechanical
properties, or the fact that it is composed of three quarks, we would have
to consider wave functions for these and to deal with the associated
equations. The resultant would be far more complex than equation (\ref{Hatom}).
Similarly, the Coulomb potential is known to be one aspect of a photon, so
that its degree of freedom has also been \emph{ignored} in this simple
equation. If the quantum mechanics of the photon is also taken into account,
we are necessarily faced with the full theory of relativistic quantum fields
\cite{Kak93}. The most venerable example of such kind of system is quantum
electrodynamics \cite{Kak93,Fey98}, in which \emph{interactions} 
between the electrons
and photons are fully incorporated. Nevertheless, by ignoring all degrees of
freedom except that associated with the electronic position, a simple 
linear equation such as (\ref{Hatom}) has been used to predict many
properties of the hydrogen atom, to a high degree of accuracy.
Parenthetically, notice that the spin degree of freedom, a favorite
in quantum computing community, is also ignored in equation (\ref{Hatom}).
When this is included into a more complex version of equation (\ref{Hatom}),
the effects of ``spin-orbit'' coupling can be discussed and a better
approximation of the properties of hydrogen emerges.

Between the simplest levels of approximation, such as equation (\ref{Hatom}%
), and the most complete/complex theories lies a vast set of approaches.
Instead of simply ignoring some of the other degrees of freedom,
these ``intermediate'' theories incorporate some of their effects into ``%
effective interactions.'' The results involve, typically, 
nonlinear equations. We give two examples here.

An early example, predating quantum electrodynamics, takes into
account the ``self interaction'' of the electron in, say, the hydrogen atom.
The idea is that, since quantum mechanical description of the electron is in
terms of a probability distribution, $P(\mathbf{x})=\left| \psi (\mathbf{x}%
)\right| ^2$ (where we have dropped the $t$ dependence for simplicity), we
are faced with a charge distribution associated with this electron: $%
\rho (\mathbf{x})=-eP(\mathbf{x})$, i.e., 
\begin{equation}
\rho (\mathbf{x})=-e\left| \psi (\mathbf{x})\right| ^2.  \label{rho}
\end{equation}
By the known laws of electrodynamics \cite{Jac98}, this charged cloud would
generate an electric potential, at every point in space, $\mathbf{x,}$ of
the form 
\begin{equation}
V_\rho (\mathbf{x})=\int \frac{\rho (\mathbf{x}^{\prime })}{\left| \mathbf{%
x-x}^{\prime }\right| }d\mathbf{x}^{\prime }.  \label{Ve}
\end{equation}
But then this potential should affect the electron in a way no different
than the potential due to the proton 
(that is, the $e^2/r$ term in Equation (\ref
{Hatom})). Inserting this extra potential into (\ref{Hatom}), we have

\begin{equation}
i\hbar \frac{\partial \psi (\mathbf{x},t)}{\partial t}=\left[ -\frac{\hbar ^2%
}{2m}\nabla ^2-\frac{e^2}r-eV_\rho (\mathbf{x})\right] \psi (\mathbf{x},t)
\label{Hatom1}
\end{equation}
Combining Equations (\ref{rho},\ref{Ve}, and \ref{Hatom1}) 
we arrive at a Schr\"{o}dinger's
equation with cubic nonlinearity:

\begin{equation}
i\hbar \frac{\partial \psi (\mathbf{x},t)}{\partial t}=\left[ -\frac{\hbar ^2%
}{2m}\nabla ^2-\frac{e^2}r\right] \psi (\mathbf{x},t)+e^2\int \frac{\psi (%
\mathbf{x},t)\left| \psi (\mathbf{x}^{\prime },t)\right| ^2}{\left| \mathbf{%
x-x}^{\prime }\right| }d\mathbf{x}^{\prime }.  \label{NLSE1}
\end{equation}
Notice that the last term describes the electron's interaction with \emph{%
itself}, as the source of the potential is the electronic cloud. We should
emphasize that this equation was eventually abandoned for a number of
reasons. One of them is that, as an approximation which keeps only
the electronic degree of freedom ($\psi $) while ignoring the photonic
degree of freedom, it is too crude to describe self interaction adequately.
As mentioned above, quantum electrodynamics was developed and the final
picture is not simply embodied in equation (\ref{NLSE1}).

Another example, which is similar in form, comes from the study of
structure of atoms with more than one electron. Instead of self
interactions, the interest here is the \emph{mutual} interactions between
the electrons. Known as the Hartree approximation, with the photonic degrees
of freedom still ignored, a system of nonlinear Schr\"{o}dinger
equations are used, even today (see Schiff \cite{Sch68}, Section 47):

\begin{equation}
i\hbar \frac \partial {\partial t}\psi _k\left( \mathbf{x}%
_k,t\right) =\left[ -\frac \hbar {2m}\nabla _k^2-\frac{Ze^2}{\left| \mathbf{%
x}_k\right| }+\sum_{j\neq k}\int |\psi _j(\mathbf{x}_j)|^2\frac{e^2}{r_{jk}}d%
\mathbf{x}_j\right] \psi _k(\mathbf{x}_k,t)\;,  \label{eqn:inter}
\end{equation}
Here, $\psi _k(\mathbf{x}_k,t)$ is the wave function of the $k^{th}$
electron in an atom with $Z$ electrons (and $Z$ protons in the nucleus) and $%
r_{jk}=|\mathbf{x}_j-\mathbf{x}_k|$ is the distance between it and the $%
j^{th}$ electron. Notice that this is a system of $Z$ nonlinear 
integrodifferential equations for the $Z$ unknowns 
$\psi _k(\mathbf{x}_k,t)$. 

Both of these examples illustrate the use of nonlinear 
Schr\"{o}dinger equations (up to cubic terms) in the context of quantum
mechanical systems. Very similar equations are also used in the context of 
nonlinear optics. In fact, many of these are even closer in form to
the ones we proposed, for example, 
\begin{equation}
\frac \partial {\partial t}\mathcal{A}(t)=c_1\mathcal{A}+c_3|\mathcal{A}|^2%
\mathcal{A} \;,  \label{NLSE2}
\end{equation}
where $c_1$ and $c_3$ are constants (\cite{Boy91}, page 280). 
Similarly, nonlinear
Schr\"{o}dinger equations of this type appear frequently in the study of
solitons \cite{Jac91}. Finally, known as the time dependent Landau-Ginzburg
equation, nonlinear systems of the form of (\ref{NLSE2}) are ubiquitous in
many areas of condensed matter physics.

In elementary quantum mechanics, the equations of evolution tend to be
linear in the wave function, describing various degrees of freedom (such as
spins) subjected to \emph{external} potentials. However, as soon as \emph{%
internal} interactions between the degrees of freedom we wish to describe
are incorporated, nonlinearities are inevitable. This section serves to
illustrate that evolution of many physical systems are governed by 
nonlinear equations. Since the equations governing the $D$ gate are
sufficiently similar to many of those in physical systems, we believe that
its implementation should be possible.


\section{Research Directions}
\label{section:future}

This paper initiates research into what problems can be solved using only 
polylogarithmic ($\log^{O(1)}n$)
entangled qubits and at most polylogarithmic steps. 
All the efforts so
far have concentrated on what can be achieved using a polynomial
number of qubits.
We have shown that constant depth QNNs of logarithmic size
have the same computational power as threshold circuits of
polynomial size and constant depth, which are used for
modeling neural networks. 
QNNs possess several advantages over threshold circuits. 
First, we need only $O(\log n)$ qubits. 
Second, there is no communication bottleneck and synchronization problems
associated with computing the weighted sum in a threshold gate;
there is no need to explicitly wire the entangled qubits together
and the synchronization is instantaneous for the computation
of the weighted sum. 
Finally, quantum systems have the ability to compute
on probability distributions rather than just discrete values, giving
them the ability to handle fuzzy sets \cite{Zad65}.

The research suggests several interesting directions.

\begin{enumerate}
\item Define a continuous version of QNNs using probability
amplitude distributions. 
Quantum 
mechanics uses infinite-dimensional Hilbert spaces to model reality.
(Even a simple system such as an electron around a nucleus in
a hydrogen atom is modeled using infinite-dimensional Hilbert space.)
How can this be used to generalize the discrete model discussed
in this paper? What kind of gates/operators are allowed? Just as we
used discrete values to encode boolean values, we can use
the continuous distribution to encode membership in fuzzy sets \cite{Zad65}.

\item How can error-correcting codes and fault-tolerant computing
be adapted to the new model, both for the discrete and continuous
versions?

\item
Following the road map provided by the theory of  neural
networks, define a theory of $\QNN(\log^{O(1)} n,
\log^{O(1)} n)$. What is the equivalent of
back-propagation and various other supervised and unsupervised
learning algorithms?  What is the equivalent of the Hopfield model and
recurrent neural networks? How can we define the formal statistical
mechanics of QNNs?


\sloppy

\item Matrix operations provide the
``killer application'' for the new model. Improve the bounds on our
results: $\TC^0 \subseteq \QNN(O(\log n),O(1))$ and $\NC \subseteq
\QNN(\log^{O(1)} n,\log^{O(1)} n)$.  In particular, minimizing the depth
of the circuits can provide real-time solutions to several important
matrix problems. Reducing the depth of the circuits to a constant,
even at the expense of increasing the size, can provide real-time 
solutions to problems in POLYLOGSPACE.
The results in this paper do not take advantage of either 
complex probability amplitudes as weights, or arbitrary unitary operators.
(This paper uses banded unitary matrices with real weights to perform 
simulations.)
Also, can we get better bounds under various uniformity constraints?

Of course, with polylogarithmic size and precision, the quantum
system has more than a polynomial number of states, and hence, it
may be infeasible to set-up the required apparatus in reasonable 
time. 
Thus, such computational systems may be more appropriate in the 
context of learning systems, where the parameters are incrementally
modified based on how the actual output of the system differs from the
required output.

\item Understand the physics behind the implementation of the new operators.
What other operators are allowed for the discrete as well as the
continuous model of QNNs? How can they be physically realized?
For example, we can define ``generalized $D$ gates'' that behave
on all the states $|i\rangle, 1\leq i\leq 2^m-1$,
as the $D$ gate behaves on the state $|0^m\rangle$ using similar 
evolutionary equation.

\item Develop a theory of $\EC$ circuits. 
While threshold circuits have been studied extensively and the
computational power of \EC\ circuits is essentially the same as the power
of threshold circuits (with polynomial size), it is not at all clear
that the \emph{learning strategies} are essentially the same. First,
\EC\ circuits can be naturally defined over the domain of complex
numbers.  We need to explore how this larger domain helps.  Second,
when translated to QNNs, \EC\ circuits must have normalized weights,
which means that the learning algorithms such as back-propagation that
make local changes to weights, cannot be directly translated without
some form of global strategy. Third, for every gate in an \EC\ circuit,
QNNs have weight vectors orthogonal to the original one.  In this
paper, these vectors simply dissipate after the $D$ operator and sink
gates. However, they may indeed be helpful in obtaining better
learning algorithms. Thus, investigating \EC\ circuits with normalized
complex weights and orthogonal weight vectors
is important, independent of their connection to
quantum computing.

\item 
Compare QNNs with the existing data in neuroscience to improve our
understanding of biological neurons and, if necessary, to refine our
model.  
It is well-known that all creatures with nervous
systems have essentially the same basic processing unit, the
biological neuron, though different types of neurons have
been identified possessing the same general structure.
The dendrites on the cell body receive inputs from several neurons,
the signal is processed in the cell body and a long axon, and is output on
the other side.
Reaction time of the lower brain for many
non-trivial tasks is less than $100$ milliseconds, while the synapse
response time of most neurons is at least $5$ milliseconds. Thus,
biological evidence suggests that the depth of these circuits cannot
exceed $20$.

It is well-known that the synaptic connections change as the brain
learns. However, there is almost no experimental evidence of a neuron
changing from an excitatory to inhibitory or vice versa as the learning
occurs. In fact, this is one of the factors cited against artificial
neural networks (threshold circuits)
and the associated learning algorithms where the
weights associated with synaptic connections can change from positive
to negative.

Thus, it is possible that the unitary operator $U$ is associated with
the synapses while the dissipation operator $D$ is associated with the
cell and the axon. Hence, the synapses are not computing on classical
bits associated with the input but the quantum interactions between
the neurotransmitters received at the input. Note that the reality is
more complicated than this simplistic model. It has been observed that
the neurons fire a series of spikes and information may be encoded in
the rate and timing of firing as well \cite{MB99}.  Of course, if we allow
feedback in QNNs, the timing issues must be considered.  The
possibility of the information being encoded in the interactions of
these carriers needs to be be investigated.

\end{enumerate}


\noindent
\textbf{Acknowledgements:} The author is indebted to Harald Hempel for 
carefully reading the first draft of the paper and suggesting numerous
improvements.


\Base

\end{document}